\documentclass[submission,copyright,creativecommons]{eptcs}
\providecommand{\event}{AFL 2017} 
\usepackage{underscore}           

\usepackage{mathrsfs}
\usepackage{amssymb}
\usepackage{amsmath}

\newcommand{\cent}{\mbox{\rm \rlap{\hskip -0.3em\raisebox{-0.15ex}{
\unitlength 1ex \linethickness{0.3pt}
\begin{picture}(0,0.0)
\put(0,0){\line(0,1){1.3}}
\end{picture}
}}c}}

\newcommand{\R}{\mbox{\sf R}}
\newcommand{\RC}{\mbox{\sf RC}}
\newcommand{\RR}{\mbox{\sf RR}}
\newcommand{\RRC}{\mbox{\sf RRC}}
\newcommand{\RW}{\mbox{\sf RW}}
\newcommand{\RRW}{\mbox{\sf RRW}}

\newcommand{\RRWW}{\mbox{\sf RRWW}}
\newcommand{\RLWW}{\mbox{\sf RLWW}}
\newcommand{\RLW}{\mbox{\sf RLW}}
\newcommand{\RL}{\mbox{\sf RL}}
\newcommand{\RLC}{\mbox{\sf RLC}}
\newcommand{\RLWWC}{\mbox{\sf RLWWC}}
\newcommand{\RRWWC}{\mbox{\sf RRWWC}}

\newcommand{\RLWWD}{\mbox{\sf RLWWD}}
\newcommand{\RRWWD}{\mbox{\sf RRWWD}}
\newcommand{\hRWW}{\mbox{\sf h-RWW}}
\newcommand{\hRRWW}{\mbox{\sf h-RRWW}}
\newcommand{\hRLWW}{\mbox{\sf h-RLWW}}
\newcommand{\hRWWC}{\mbox{\sf h-RWWC}}
\newcommand{\hRRWWC}{\mbox{\sf h-RRWWC}}
\newcommand{\hRLWWC}{\mbox{\sf h-RLWWC}}
\newcommand{\hRWWD}{\mbox{\sf h-RWWD}}
\newcommand{\hRRWWD}{\mbox{\sf h-RRWWD}}
\newcommand{\hRLWWD}{\mbox{\sf h-RLWWD}}

\newcommand{\beginproof}{{\noindent \bf Proof.~}}
\newcommand{\myendproof}{\hspace*{\fill} $\Box$ \vspace{+0.2cm}}

\newcommand{\lang}[1]{{\cal L}(#1)}

\newcommand{\REG}{\makebox{\sf REG}}
\newcommand{\CFL}{\makebox{\sf CFL}}

\newcommand{\DCFL}{\makebox{\sf DCFL}}

\newcommand{\MVR}{\makebox{\sf MVR}}

\newcommand{\SL}{\makebox{\sf SL}}

\newcommand{\Restart}{\makebox{\sf Restart}}
\newcommand{\Accept}{\makebox{\sf Accept}}
\newcommand{\Reject}{\makebox{\sf Reject}}

\newcommand{\calL}[1]{{\cal L}(#1)}
\newcommand{\cL}{{\cal L}}
\newcommand{\N}{\mathbb{N}}
\newcommand{\calLP}[1]{{\cal L}_{hP}(#1)}

\newtheorem{lemma}{Lemma}{\bf}{\it}
\newtheorem{proposition}[lemma]{Proposition}{\bf}{\it}
\newtheorem{theorem}[lemma]{Theorem}{\bf}{\it}
\newtheorem{corollary}[lemma]{Corollary}{\bf}{\it}
\newtheorem{definition}[lemma]{Definition}{\bf}{\it}
{\bf}{\rm}
{\bf}{\rm}
\newtheorem{fact}[lemma]{Fact}{\bf}{\it}

\title{On h-Lexicalized Restarting Automata}
\author{Martin Pl\'atek
\institute{Charles University\\
Department of Computer Science\\
Malostransk\'e n\'am.~25\\
118 00 PRAHA, Czech Republic}
\email{martin.platek@mff.cuni.cz}
\and
Friedrich Otto
\institute{Universit\"at Kassel\\
Fachbereich Elektrotechnik/Informatik\\
34109 KASSEL, Germany}
\email{\quad otto@theory.informatik.uni-kassel.de}
}

\def\titlerunning{On h-Lexicalized Restarting Automata}
\def\authorrunning{M.~Pl\'atek \& F.~Otto}
\begin{document}
\maketitle

\begin{abstract}
Following some previous studies on restarting automata,
we introduce a refined  model --
the \emph{h-lexicalized restarting automaton} (\hRLWW).
We argue that this model is useful for expressing lexicalized syntax in computational linguistics.
We compare the input languages, which are the languages traditionally considered in automata theory,
to the so-called \emph{basic} and \emph{h-proper languages},
which are (implicitly) used by categorial grammars, the original tool for the description of lexicalized syntax.
The basic and h-proper languages allow us to stress several nice properties of h-lexicalized restarting automata,
and they are suitable for modeling the analysis by reduction
and, subsequently, for  the development of categories of a lexicalized syntax.
Based on the fact that
a two-way deterministic monotone restarting automaton can be transformed into an equivalent
deterministic monotone RL-automaton in (Marcus) contextual form,
we obtain a transformation from monotone RLWW-automata that recognize  the class CFL of context-free languages
as their input languages  to deterministic monotone h-RLWW-automata that recognize CFL through their h-proper languages.
Through this transformation we obtain automata with the \emph{complete correctness preserving property}
and an infinite hierarchy within CFL, based on the size of the read/write window.
Additionally, we consider h-RLWW-automata that are allowed to perform multiple rewrite steps per cycle,
and we establish another infinite hierarchy above CFL that is based on the number of rewrite steps that
may be executed within a cycle.
The corresponding separation results and their proofs illustrate the transparency of h-RLWW-automata
that work with the (complete or cyclic) correctness preserving property.
\end{abstract}

\section{Introduction}\label{s1}
The linguistic technique of `analysis by reduction' is used to analyze sentences of natural
languages with a high degree of word-order freedom like, e.g., Czech, Latin, or German~(see, e.g.,~\cite{LPK05}).
A human reader is supposed to understand the meaning of a given sentence
before he starts to analyze it.
Analysis by reduction (partially) simulates such a behavior
by analyzing sentences, where
morphological and syntactical tags have been added to the word-forms and punctuation marks~(see, e.g.,~\cite{LPS07}).

In~\cite{JMPV95} the restarting automaton was presented as a formal device to model the
naive (i.e. non-tagged)  analysis by reduction.
Such a restarting automaton has a finite-state control and a flexible tape with endmarkers
on which a window of fixed finite size operates.
The automaton works in cycles, where each cycle begins with the automaton being in its initial state
with the window over the left end of the tape.
Now it scans the tape from left to right until, at some place, it performs a combined rewrite/restart operation
that deletes one or more symbols from the window, moves the window back to the left end of the tape,
and returns the automaton to the initial state.
As the tape is flexible, it adjusts automatically to the now shortened inscription.
This model, nowadays called R-automaton, is quite restricted.
Accordingly,
in subsequent years this model was extended by allowing more general rewrite operations,
by admitting additional symbols (so-called auxiliary symbols) in the working alphabet,
and by separating the rewrite operation from the restart operation.
In addition, models were proposed that can move their window in both directions,
enabling them to first scan the given input completely before executing any rewrite step
(for a survey see~\cite{otto03} or \cite{otto06}).

To model the analysis by reduction on sequences of tagged items, \emph{basic languages} and
\emph{proper languages} of restarting automata were considered in~\cite{MOPTCS410}.
While the \emph{input language} of an automaton $M$ just consists of all input words
that $M$ can accept, the \emph{basic language} of $M$ consists of all words over the working alphabet
that $M$ can accept, and the \emph{proper language} consists of all words that are obtained from
the basic language by erasing all non-input (that is, auxiliary) symbols.
Now one can argue that the auxiliary symbols represent the tags, and so,
the basic language models the sequences of tagged items that are accepted.
However, as it turned out, there are already deterministic restarting automata for which
these proper languages are not even recursive, although the input language (and the basic language)
of each deterministic restarting automaton is decidable in polynomial time.

Hence, \emph{lexicalized} types of restarting automata were introduced in~\cite{MOPTCS410},
in which the use of auxiliary symbols is somewhat restricted.
A lexicalized restarting automaton is deterministic and there is a positive constant
$c$ such that essentially each factor of a word from the basic language that only consists of auxiliary
symbols has length at most~$c$.
One of the main results of~\cite{MOPTCS410} states that the class of proper languages of lexicalized
monotone RRWW-automata (see Sections~\ref{s2} and~\ref{s5} for the definitions) coincides with the class
of context-free languages.

Here, in order to give a theoretical basis for lexicalized syntax,
we introduce a  model of the restarting automaton
that formalizes lexicalization in a similar way as categorial grammars (see e.g.~\cite{Bar53})
-- the \emph{h-lexicalized restarting automaton} ({\hRLWW}).
This model is obtained from the two-way restarting automaton of~\cite{Pla01}
by adding a letter-to-letter morphism $h$ that assigns an input symbol to each working symbol.
Then the \emph{h-proper language} of an h-RLWW-automaton $M$
consists of all words $h(w)$, where $w$ is taken from the basic language of~$M$.
Thus, in this setting the auxiliary symbols themselves play the role of the tagged items,
that is, each auxiliary symbol $b$ can be seen as a pair consisting of an input symbol $h(b)$ and
some additional information (tags).
We  argue that this new model is better suited to the
modeling of the lexicalized syntactic analysis and (lexicalized) analysis by reduction of natural languages
(compare \cite{LPK05,LPS07})
through the use of basic and h-proper languages.
We stress the fact that this model works directly with the text-editing operations (rewrite, delete).

We recall some constraints that are typical for restarting automata,
and we outline ways for new combinations of constraints.
As our basic technical result,
we show that the expressive power of  two-way deterministic  monotone restarting automata
(det-mon-RLWW-automata) does not decrease,
if we use the corresponding type of automaton which instead of rewritings can only delete symbols (det-mon-RL-automata)
and which are in the so-called (Marcus) \emph{contextual form} (for \emph{contextual grammars} see \cite{Mar69,HandII}).
In fact, these types of automata all characterize the class LRR of left-to-right regular languages~\cite{CuCo73}.
The technique for this transformation is derived from the linguistic techniques for dependency syntax.

Then we show that the h-proper languages of monotone h-RRWW- and h-RLWW-automata just yield the
context-free languages, both in the deterministic and in the nondeterministic case.
In particular, this means that h-proper languages of deterministic monotone h-RLWW-automata
properly extend the input languages of this type of automaton.
Then we prove that the h-proper language of any h-RLWW-automaton occurs as the input language
of some \emph{shrinking} h-RLWW-automaton,
but that there are input languages of RLWW-automata that do not occur as h-proper languages of h-RLWW-automata.
Finally, based on the size of the read/write window, we establish infinite ascending hierarchies
of language classes of h-proper languages for several subtypes of h-RLWW-automata.
In particular, we obtain hier\-archies within LRR and within CFL that have LRR and CFL as their limits.
In addition, we also study h-RLWW-automata that are allowed to execute several rewrite steps within a cycle,
and we establish infinite ascending hierarchies above CFL that are based on the number of rewrite steps
that are executed within a cycle.

The paper is structured as follows.
In Section~\ref{s2}, we introduce our model and its submodels,
we define the h-proper languages, and we state the complete correctness preserving property
and the complete error preserving property for the basic languages of deterministic h-RLWW-automata.
In Section~\ref{s5}, we present the aforementioned characterization of the context-free languages
through h-proper languages of deterministic monotone h-RLWW-automata,
and we discuss the relationship between input and h-proper languages for some types
of h-RLWW-automata.
Then, in Section~\ref{seclookahead}, we derive the aforementioned hierarchies
inside LRR and CFL, and in Section~\ref{Secmultp} we consider h-RLWW-automata with several rewrites per cycle.
The paper concludes with Section~\ref{SecConclusion}
in which we summarize our results and state some problems for future work.

\section{Definitions}\label{s2}
We start with the  definition of the two-way restarting automaton.
For technical  reasons we propose a slight modification
from the original definition given in~\cite{Pla01}.

\begin{definition}\label{DefRLWW}
A \emph{two-way restarting automaton}, an \emph{RLWW-automaton} for short,
is a machine with a single flexible tape and a finite-state control.
It is defined through
an 8-tuple
$M=(Q,\Sigma,\Gamma,\cent,\$,q_0,k,\delta)$,
where $Q$ is a finite set of states,
$\Sigma$ is a finite input alphabet, and
$\Gamma$ is a finite working alphabet containing~$\Sigma$.
The symbols from $\Gamma\smallsetminus \Sigma$ are called \emph{auxiliary symbols}.
Further,
the symbols $\cent,\$ \not\in\Gamma$, called \emph{sentinels}, are the markers for the left and right
border of the workspace, respectively,
$q_0\in Q$ is the initial state,
$k\ge 1$ is the size of the \emph{read/write window}, and
\begin{center}
$\delta:Q\times {\mathcal {PC}}^{\le k} \to {\mathcal P}((Q\times\{{\sf MVR}, {\sf MVL}, \mbox{\sf SL($v$)}\})
\cup \{{\sf Restart},{\sf Accept}, {\sf Reject}\})$
\end{center}
is the \emph{transition relation}.
Here ${\mathcal P}(S)$ denotes
the powerset of a set~$S$, 
$${\mathcal {PC}}^{\le k} = (\cent\cdot \Gamma^{k-1}) \cup \Gamma^k \cup
(\Gamma^{\le k-1}\cdot\$) \cup (\cent\cdot\Gamma^{\le k-2} \cdot
\$)$$
is the set of \emph{possible contents} of the read/write window of~$M$,
and  $v\in {\mathcal {PC}}^{\le k-1}$.
\vspace{+0.2cm}

\noindent
The transition relation describes six different types of
transition steps (or instructions):
\begin{itemize}
\item[1.] A \emph{move-right step} $(q,u)\longrightarrow (q', {\sf MVR})$  assumes that  $(q',{\sf MVR})\in \delta(q,u)$,
where $q,q'\in Q$ and
$u\in{\cal PC}^{\le k}$,
$u\not=\$ $.
If $M$ is in state $q$ and sees the word $u$ in its
read/write window,
then this move-right step causes $M$ to shift the
window one position to the right
and to enter state~$q'$.

\item[2.] A \emph{move-left step} $(q,u)\longrightarrow (q', {\sf MVL})$  assumes that
$(q',{\sf MVL})\in \delta(q,u)$,
where $q,q'\in Q$ and
$u\in{\cal PC}^{\le k}$,
$u\not\in\cent\cdot \Gamma^*\cdot\{\lambda,\$\}$
(Here $\lambda$ is used to denote the \emph{empty word}).
It causes $M$ to shift the 
window one position to the left
and to enter state~$q'$.

\item[3.] An \emph{SL-step}
$(q,u)\longrightarrow (q', \mbox{\sf SL$(v)$})$
assumes that $(q',\mbox{\sf SL$(v)$})\in \delta(q,u)$,
where $q,q'\in Q$,  $u\in{\cal PC}^{\le k}$, and $v\in{\cal PC}^{\le k-1}$,
that $v$ is  shorter than~$u$,
and that $v$ contains all the sentinels that occur in~$u$ (if any).
It causes $M$ to replace $u$ by~$v$, to enter state~$q'$, and to shift the
window
by $|u|-|v|$ items to the left --
but at most to the left sentinel~$\cent$
(that is, the contents of the window is `completed' from the left,
and so the distance to the left sentinel decreases if the window
was not already at~$\cent$).

\item[4.] A \emph{restart step}
$(q,u)\longrightarrow {\sf Restart}$ assumes that
${\sf
Restart} \in \delta(q,u)$, where $q\in Q$ and $u\in{\cal PC}^{\le k}$.
It causes $M$ to move its
window to the left end
of the tape, so that the first symbol it sees is the left
sentinel~$\cent$, and to reenter the initial state~$q_0$.

\item[5.]
An \emph{accept step}
$(q,u)\longrightarrow {\sf Accept}$ assumes that
${\sf Accept} \in \delta(q,u)$, where $q\in Q$ and $u\in{\cal PC}^{\le k}$.
It causes $M$ to halt and accept.

\item[6.] A \emph{reject step}
$(q,u)\longrightarrow {\sf Reject}$ assumes that
${\sf Reject} \in \delta(q,u)$, where $q\in Q$ and $u\in{\cal PC}^{\le k}$.
It causes $M$ to halt and reject.
\end{itemize}
\end{definition}

There are two differences to the original definition given in~\cite{Pla01}
in that we have explicit reject steps 
and in that after a rewrite, that is, an \SL-step, the window is not moved but just refilled from the left.
It is easily seen that these modifications do not influence the expressive power of the model.

A \emph{configuration} of an RLWW-automaton $M$ is a word $\alpha q\beta$,
where $q\in Q$, and either $\alpha=\lambda$ 
and
$\beta\in\{\cent\}\cdot\Gamma^*\cdot\{\$\}$
or $\alpha\in \{\cent\}\cdot\Gamma^*$ and
$\beta\in \Gamma^*\cdot\{\$\}$;
here $q$ represents the current state,
$\alpha\beta$  is the current contents of the tape, and
it is understood that
the read/write window contains the first $k$ symbols of $\beta$ or
all of $\beta$ if
$|\beta|<k$.
A \emph{restarting configuration}
is of the form $q_0\cent w\$$, where $w\in\Gamma^*$;
if $w\in \Sigma^*$, then $q_0\cent w\$$
is an \emph{initial configuration}.
We see that any initial configuration is also a restarting configuration, and that
any restart transfers $M$ into a restarting configuration.

In general, an RLWW-automaton $M$ is \emph{nondeterministic}, that is,
there can be two or more steps (instructions) with the same left-hand side $(q,u)$,
and thus, there can be more than one computation that start from a given restarting configuration.
If this is not the case, the automaton is \emph{deterministic}.

A \emph{computation} of $M$ is a sequence $C = C_0,C_1,\ldots,C_j$ of configurations of~$M$, where
$C_0$ is an initial or a restarting configuration and $C_{i+1}$ is obtained from $C_i$ by a step of~$M$,
for all $0 \le i <j$.
In the following we only consider computations of RLWW-automata
which end either by an accept or by a reject step.
\vspace{+0.1cm}

\noindent
{\bf -- Cycles and tails:}
Any finite computation of an  RLWW-automaton $M$ consists of certain phases.
A phase, called a \emph{cycle},
starts in a restarting configuration,
the window moves along the tape performing non-restarting steps 
until a {restart} step is
performed and thus a new restarting configuration is reached.
If no further restart step is performed, any finite computation
necessarily finishes in a halting configuration -- such a phase is
called a \emph{tail}.
Here we require that $M$ executes \emph{exactly one} SL-step within any cycle,
and that it does not execute any  SL-step in a tail.
The latter is again a deviation from the original model, but in the nondeterministic setting,
this does not influence the expressive power of the model, either.

We use the notation $q_0\cent u\$ \vdash_M^c q_0\cent v\$$ to denote a cycle of $M$
that begins with the restarting configuration $q_0\cent u\$$ and ends with the
restarting configuration $q_0\cent v\$$.
Through this relation we define the relation of \emph{cycle-rewriting} by~$M$.
We write $u \Rightarrow_M^c  v$ iff $q_0\cent u\$ \vdash_M^c q_0\cent v\$$ holds.
The relation  $  \Rightarrow_M^{c^*}  $ is
the reflexive and transitive closure of $  \Rightarrow_M^c $.
We stress that the cycle-rewriting is a very important feature of an RLWW-automaton.
As each SL-step is strictly length-reducing, we see that
$u \Rightarrow_M^{c}  v$ implies that  $|u| > |v|$.
Accordingly,
$u \Rightarrow_M^{c}  v$  is also called a \emph{reduction} by~$M$.
\vskip 1mm

An  \emph{input  word $w\in \Sigma^*$ is accepted by $M$}, if there
is a computation which starts with the initial configuration
$q_0\cent w\$$ and ends by executing an accept step.
By $L(M)$ we denote the language consisting of all input words accepted
by $M$; we say that \emph{$M$ recognizes (or accepts) the input language
$L(M)$}.
\vskip 1mm

A \emph{basic (or characteristic)  word} $w\in \Gamma^*$ is accepted by $M$
if there
is a computation which starts with the restarting configuration
$q_0\cent w\$$ and ends by executing an accept step.
By $L_C(M)$ we denote the language consisting of all words from $\Gamma^*$ that are accepted
by $M$; we say that \emph{$M$ recognizes (or accepts) the basic (or characteristic)  language
$L_C(M)$}.
\vskip1mm

By ${\mathrm Pr}^\Sigma$ we denote the projection from $\Gamma^*$ onto~$\Sigma^*$,
that is, ${\mathrm Pr}^\Sigma$ is the morphism defined by $a\mapsto a$ for all $a\in\Sigma$
and $A\mapsto \lambda$ for all $A\in\Gamma\smallsetminus \Sigma$.
If $v={\mathrm Pr}^\Sigma(w)$, then $v$ is the \emph{$\Sigma$-projection} of~$w$,
and $w$ is an \emph{extended version} of~$v$.
The \emph{proper language} of $M$ is now the language $L_P(M)={\mathrm Pr}^\Sigma(L_C(M))$,
that is, a word $v\in\Sigma^*$ belongs to $L_P(M)$ iff there exists an expanded version
$w$ of $v$ such that $w\in L_C(M)$.

Finally, we come to the definition of the central notion of this paper, the h-lexicalized RLWW-automaton.

\begin{definition}
An \emph{h-lexicalized RLWW-automaton}, or {\sf h-RLWW}-automaton, is a pair $\hat{M}=(M,h)$, where
$M=(Q,\Sigma,\Gamma,\cent,\$,q_0,k,\delta)$ is  an RLWW-automaton and
$h:\Gamma\to\Sigma$ is a letter-to-letter morphism satisfying $h(a) = a$ for all input letters $a\in\Sigma$.
The input language $L(\hat{M})$ of $\hat{M}$ is simply the language $L(M)$ 
and the basic language $L_C(\hat{M})$ of $\hat{M}$
is the language $L_C(M)$.
Finally, we take $L_{hP}(\hat{M}) = h(L_C(M))$,
and we say that \emph{$\hat{M}$ recognizes (or accepts) the h-proper language
$L_{hP}(\hat{M})$}.
\end{definition}

\noindent
\emph{Notation.} For brevity, the prefix {\sf det-} will
be used to denote the property of being deterministic.
For any type {\sf A} of restarting automaton, $\lang{\mbox{\sf A}}$ will denote the
class of input languages that are recognized by automata  from~{\sf A},
${\mathcal L}_C({\mbox{\sf A}})$ will denote the class of basic languages that are recognized by automata from~{\sf A},
and ${\mathcal L}_{hP}({\mbox{\sf A}})$ will denote the class of h-proper languages that are recognized by automata from~{\sf A}.
For a natural number $k\ge 1$, $\lang{\mbox{\sf $k$-A}}$
(${\mathcal L}_C({\mbox{\sf $k$-A}})$ or ${\mathcal L}_{hP}({\mbox{\sf $k$-A}})$) will denote
the class of input (basic or h-proper) languages that are recognized by those automata
from {\sf A} that use a read/write window of size~$k$.

By $\subset$ we denote the proper subset relation.
Finally,
$\N_+$ will denote the set of all positive integers.

\subsection{Further refinements and constraints on (h-lexicalized) RLWW-automata}\label{s4}
Here we introduce some constrained types of rewrite steps.

A \emph{delete-left step}
$(q,u)\rightarrow (q', \mbox{\sf DL($v$)})$
assumes that $(q',\mbox{\sf SL($v$)}) \in \delta(q,u)$ and that
$v$ is a proper subsequence of~$u$, containing
all the sentinels from~$u$ (if any).
It causes $M$ to replace $u$ by $v$ (by deleting excessive symbols),
to enter state~$q'$, and to shift the window
by $|u|-|v|$ symbols to the left,
but at most to the left sentinel~$\cent$.
Hence, the contents of the window is `completed' from the left,
and so the distance
of the window to the left sentinel decreases
if the window was not already at the left sentinel.

A \emph{contextual-left step}
$(q,u)\rightarrow (q', \mbox{\sf CL($v$)})$
assumes that $(q',\mbox{\sf SL($v$)}) \in \delta(q,u)$,
where $u = v_1u_1v_2u_2v_3$ and $v= v_1v_2v_3$
such that $v$ contains all the sentinels from~$u$ (if any).
It causes $M$ to replace $u$ by~$v$ (by deleting the factors $u_1$ and $u_2$ of~$u$),
to enter state~$q'$, and to shift the window by $|u|-|v|$ symbols to the left,
but at most to the left sentinel~$\cent$.
\vskip 1mm

An RLWW-automaton is an \emph{RLW-automaton} if its working
alphabet coincides with its input alphabet, that is,
no auxiliary symbols are available to this automaton.
Note that in this situation, each restarting configuration is necessarily an initial configuration.

An RLW-automaton is an \emph{RL-automaton} if all its rewrite steps are {DL}-steps,
and it is an \emph{RLC}-automaton if all its rewrite steps are CL-steps.
Further, an RLWW-automaton is an \emph{{RLWWC}-automaton} 
(that is, an \emph{RLWW-automaton in Marcus contextual form})
if all its rewrite steps are {CL}-steps.
Similarly, an RLWW-automaton is an \emph{{RLWWD}-automaton}
if all its rewrite steps are {DL}-steps.
Observe that when concentrating on input languages, then DL- and CL-steps  ensure
that no auxiliary symbols can ever occur on the tape; if, however,
we are interested in basic or h-proper languages,
then auxiliary symbols can play an important role even though a given RLWW-automaton
uses only DL- or CL-steps.
Therefore, we distinguish between RLWWC- and RLC-automata, and between RLWWD- and RL-automata.

An RLWW-automaton is an \emph{RRWW-automaton} if it does not use any {MVL}-steps.
{}From these automata, we obtain \emph{RRW-}, \emph{RR-}, \emph{{RRC}-},
\emph{{RRWWD}-}, and \emph{{RRWWC}-automata} in an analogous way.
Further, an RRWW-automaton is an \emph{RWW-automaton} if it restarts
immediately after executing an SL-step. 
{}From these automata, we obtain
\emph{RW-},  \emph{R-}, \emph{{RC}-},  \emph{{RWWD}-},
and \emph{{RWWC}-automata}.
Obviously, from these types of automata, we obtain corresponding types of h-lexicalized automata,
denoted by the prefix {h-}.
Observe that for an h-RLW-automaton $(M,h)$,
the letter-to-letter morphism $h$ is necessarily the identity mapping on~$\Sigma$.
Hence, for such an automaton, the input language, the basic language, and the h-proper language all coincide.

We have the following simple facts, which illustrate the
transparency of computations of deterministic RLWW-automata with respect to their basic languages.

\begin{fact}\label{factCPP}
{\bf (Complete Correctness Preserving Property for ${\mathcal L}_C(\mbox{\sf det-RLWW})$)}\\
Let $M$ be a deterministic RLWW-automaton, let
$C = C_0, C_1, \ldots, C_n$ be a computation of $M$,
and let $\cent u_i\$$ be the tape contents of the configuration~$C_i$, $0\le i\le n$.
If $u_i\in L_C(M)$ for some~$i$, then $u_j\in L_C(M)$ for all $j=0,1,\ldots,n$.
\end{fact}

\begin{fact}\label{factEPP}
{\bf (Complete Error Preserving Property for ${\mathcal L}_C(\mbox{\sf det-RLWW})$)}\\
Let $M$ be a deterministic RLWW-automaton,
let $C = C_0, C_1, \ldots, C_n$ be a computation of $M$,
and let $\cent u_i\$$ be the tape contents of the configuration~$C_i$, $0\le i\le n$.
If $u_i\not\in L_C(M)$ for some~$i$, then $u_j\not\in L_C(M)$ for all $j=0,1,\ldots,n$.
\end{fact}

\begin{fact}\label{factEPPnd}
{\bf (Error Preserving Property for ${\mathcal L}_C(\mbox{\sf RLWW})$)}\\
Let $M$ be an RLWW-automaton,
let $C = C_0, C_1, \ldots, C_n$ be a computation of~$M$,
and let $\cent u_i\$$ be the tape contents of the configuration~$C_i$, $0\le i\le n$.
If $u_i \notin L_C(M)$ for some~$i$, then
$u_j\not\in L_C(M)$ for all $j\ge i$.
\end{fact}

\subsection{Proper languages versus h-proper languages}\label{SsecProper}
As each SL-step is length-reducing, and as each cycle of each computation of an RLWW-automaton
contains an application of an SL-step,
it is easily seen that the membership problem for the basic language $L_C(M)$ is decidable
in polynomial time for each deterministic RLWW-automaton~$M$.
On the other hand,
in~\cite{MOPTCS410} it is shown that there exist deterministic RRWW-automata
for which the corresponding proper languages are non-recursive.
This fact is derived from the result that, for each recursively enumerable language
$L\subseteq \Sigma_0^+$, where $\Sigma_0 = \{ a, b\}$,
there exists a deterministic RRWW-automaton $M$ such that $L_P(M)\cap \Sigma_0^+\cdot c =\varphi(L)$,
where $c$ is an additional symbol, and $\varphi$ is a finite-state transduction.
Thus, these proper languages are much more sophisticated than the basic languages.
In particular, there is no general relation between the terminal symbols and the auxiliary symbols in
a word from the basic language~$L_C(M)$.

To remedy this, a notion of \emph{lexicalization} was introduced in~\cite{MOPTCS410}.
Let $\mathrm{NIR}(M)$ denote the set of all words from $\Gamma^*$ that are not immediately
rejected by~$M$, that is, the words that are either accepted in a tail or that cause $M$ to perform at least one cycle.
An RRWW-automaton $M$ is called \emph{lexicalized}
if it is deterministic and if there exists a positive integer constant $j$ such that,
whenever a word $v\in(\Gamma\smallsetminus\Sigma)^*$ is a factor of a word $w\in\mathrm{NIR}(M)$,
then $|v|\le j$, that is,
factors of words from $\mathrm{NIR}(M)$
that only consist of auxiliary symbols are at most of length~$j$.
It was then shown that the proper language $L_P(M)$ of any lexicalized RRWW-automaton
is growing context-sensitive.

In the current paper we use the notion of \emph{h-lexicalization} to obtain a stronger correspondence
between the auxiliary symbols and the terminal symbols in a word, as each auxiliary symbol
is mapped to a terminal symbol through the given morphism~$h$.
This corresponds to the process of annotation of a given terminal word, replacing each
given lexical item through a tuple containing this item together with morphological
and lexical {syntactic} information, which then form the basis for the subsequent analysis by reduction.

\section{Robustness of monotone RLWW-automata}\label{s5}
We recall the notion of \emph{monotonicity} as an important constraint for computations of restarting automata.
Let $M$ be an RLWW-automaton, and
let $C = C_k,C_{k+1},\ldots,C_j$ be a sequence  of configurations of~$M$,
where $C_{i+1}$ is obtained by a single transition step from~$C_i$, $k \le i < j$.
We say that $C$ is a \emph{subcomputation} of~$M$.
If $C_i=\cent\alpha q\beta\$$,
then $|\beta\$|$ is the \emph{right distance} of $C_i$,
which is denoted by $D_r(C_i)$.
We say that a subsequence
$(C_{i_1}, C_{i_2}, \ldots,C_{i_n})$ of $C$ is \emph{monotone}
if $D_r(C_{i_1}) \ge D_r(C_{i_2}) \ge  \cdots \ge D_r(C_{i_n})$.
A computation of $M$ is called \emph{monotone} if the corresponding subsequence of
rewrite configurations is monotone.
Here a configuration is  called a \emph{rewrite configuration} if in this configuration
an SL-step is being applied.
Finally, $M$ itself is called \emph{monotone} if each of its computations is monotone.
We use the prefix \emph{mon-} to denote monotone types of RLWW-automata.
This notion of monotonicity has been considered before in various papers.
The following results have been established.

\begin{theorem}\label{m2}{\rm \cite{JMPV99,JMOP05a,JOMP05,O09,Pla01}}
\vspace{0.1cm}

\noindent
$
\begin{array}{lccccccc}
{\rm (a)} & \CFL & = & \cL(\mbox{\sf mon-RWW}) & = & \cL(\mbox{\sf mon-RRWW})
& = & \cL(\mbox{\sf mon-RLWW}).\\[+0.1cm]
{\rm (b)} & {\sf LRR} & = & \cL(\mbox{\sf det-mon-RL}) & = & \cL(\mbox{\sf det-mon-RLW})
& = & \cL(\mbox{\sf det-mon-RLWW}).\\[+0.1cm]
{\rm (c)} & \DCFL & = & \cL(\mbox{\sf det-mon-RC}) &  =
& \cL(\mbox{\sf det-mon-RR}) & = &  \cL(\mbox{\sf det-mon-RRWW}).
\end{array}
$
\end{theorem}

Here DCFL denotes the class of \emph{deterministic context-free languages},
LRR is the class of \emph{left-to-right regular languages} from~\cite{CuCo73},
and CFL is the class of \emph{context-free languages}.
Actually the result in Theorem~\ref{m2} (b) can be strengthened as follows.

\begin{theorem}\label{m2b}
${\sf LRR}  =  \cL(\mbox{\sf det-mon-RLC}) = \cL(\mbox{\sf det-mon-RLWW}).$
\end{theorem}

Thus, for each det-mon-RLWW-auto\-ma\-ton~$M_a$,
there is a det-mon-RLC-automaton~$M_b$ that accepts the same input language.

\vspace{+0.1cm}
\noindent
{\bf Proof outline.}
To prove this result one
must overcome the problem that
there is no straightforward simulation of a det-mon-RLWW-automaton
by a det-mon-RLC-automaton.
This follows from the observation that any det-mon-RLC-automaton fulfills
the Complete Correctness Preserving Property for its input language
and that, moreover, all its rewrite steps are contextual.
On the other hand, the det-mon-RLWW-automaton $M_a$
will in general not satisfy any type of correctness preserving property for its input language.
This means in particular that 
the reductions of
$M_a$ and $M_b$ will in general differ substantially.
The second difficulty results from the problem of how to ensure that
$M_b$ chooses the correct places for executing CL-steps
in a deterministic fashion without the ability to use any non-input symbols.

In~\cite{POMnew} these problems are solved through a sequence of transformations.
First it is shown that each det-mon-RLWW-automaton $M_a$ can be transformed into
an equivalent automaton $M_1$ that uses additional length-preserving rewrite steps,
that has a window of size two only, and that works in three well-defined passes.
During the first pass, $M_1$ just applies a sequence of MVR-steps until it reaches the right sentinel.
During the second pass $M_1$ applies length-preserving rewrite steps and MVL-steps
until the window reaches the left sentinel again.
In this way information is encoded at each position of the input on the suffix to the right of that position.
Finally, in the third pass $M_1$ only applies length-preserving rewrite steps, MVR-, and SL-steps,
simulating the actual reductions of~$M_a$.
In addition, $M_1$ is monotone with respect to its SL-steps.
As $M_1$ has a window of size two, each of its SL-steps can be interpreted as replacing the left symbol inside the window
and deleting the right symbol from the window.
It follows that the computation of $M_1$ can be described by a tree-like graph, called an SL-graph,
that has nodes in one-to-one correspondence to the initial tape contents.
Essentially, an SL-graph describes through its (oriented) edges
which symbols are deleted with which symbols as their immediate left neighbours.
Based on some combinatorial arguments cutting lemmas can be established for these graphs.
These in turn lead to a simulation of $M_1$ by an automaton $M_2$ that
works in so-called \emph{strong cycles}, and that uses CL-steps
instead of SL-steps.
Each strong cycle of $M_2$ consists of three passes, similar to the passes of~$M_1$,
where the third pass ends with a \emph{strong restart} that resets the automaton into its initial state,
replaces each symbol currently on the tape by its associated input symbol with respect to the letter-to-letter
morphism of~$M_2$, and moves the window to the left end of the tape.
Further, the window of $M_2$ is quite large, as
the CL-steps of $M_2$ do not correspond to the SL-steps of~$M_1$,
but rather a single CL-step simulates the effect of a whole sequence of SL-steps.
Finally, it can be shown that the effect of the three passes of the strong cycles of $M_2$ and the subsequent
strong restart step can be
described through a sequence of four deterministic two-way finite-state transducers.
As the class of functions that are computable by these transducers is closed under composition~\cite{AU70},
it follows that the transformation on initial configurations that is induced by the strong cycles of $M_2$
can be realized by a deterministic two-way finite-state transducer.
Now this transducer can be
converted into a det-mon-RLC-automaton $M_b$
that still accepts the same language as~$M_a$.
\hspace*{\fill}$\Box$

\subsection{Robustness of monotonicity and h-proper languages}
The following theorem extends the basic theorem from \cite{MOPTCS410} and completes the
aforementioned characterization of the context-free languages.

\begin{theorem}\label{LPandCFL}
$\begin{array}[t]{lcccc}
\mbox{\sf CFL} & = &  {\mathcal L}_{hP}({\mbox{\sf det-mon-h-RRWWC}}) &  = &{\mathcal L}_{hP}({\mbox{\sf mon-h-RRWW}})\\
                             & = & {\mathcal L}_{hP}({\mbox{\sf det-mon-h-RLWW}}) & = & {\mathcal L}_{hP}({\mbox{\sf mon-h-RLWW}}).
\end{array}$
\end{theorem}

\beginproof
The proof is based mainly on ideas from \cite{MOPTCS410}.
Here it is carried over from monotone lexicalized RRWW-automata to monotone h-RLWW-automata.

Let $M=((Q,\Sigma,\Gamma,\cent,\$,q_0,k,\delta),h)$ be a monotone h-RLWW-automaton.
Then the characteristic language $L_{C}(M)$ is context-free~\cite{JMPV99,Pla01}.
As $L_{hP}(M) = h(L_{C}(M))$, and as CFL is closed
under morphisms, it follows that $L_{hP}(M)$ is context-free.

Conversely, assume that $L\subseteq \Sigma^*$ is a context-free language.
Without loss of generality we may assume that $L$ does not contain the empty word.
Thus, there exists a context-free grammar $G = (N,\Sigma,S,P)$ for~$L$ that is in
\emph{Greibach normal form}, that is,
each rule of $P$ has the form $A\to a\alpha$  for some symbol $a \in \Sigma$
and for some word $\alpha\in  N^{*}$ (see, e.g., \cite{HoUl79}).
For the following construction we assume
that the rules of $G$ are numbered from 1 to~$m$.

{F}rom $G$ we construct a new grammar $G'=(N,B,S,P')$,
where $B=\{\,(\nabla_i, a)\mid 1\le i\le m\,$ and the $i$-th rule of $G$ has the form $A\to a\alpha\, \}$
is a set of new terminal symbols that are in one-to-one correspondence to the rules of~$G$,
and
$P'=\{\,A\to (\nabla_i, a)\alpha\mid  A\to a\alpha \mbox{ is the }i\mbox{-th rule of }G, \;1\le i\le m\,\}.$

Obviously, a word $\omega\in B^*$
belongs to $L(G')$ if and only if $\omega$ has the form
$$\omega =
(\nabla_{i_1},a_1)(\nabla_{i_2},a_2)\cdots(\nabla_{i_n},a_n)$$
for some integer $n>0$, where $a_1,a_2,\ldots,a_n\in \Sigma$,
$i_1,i_2,\ldots,i_n\in\{1,2,\ldots,m\}$, and the sequence of  indices $(i_1,i_2,\ldots,i_n)$ describes a
(left-most) derivation of
$w=a_1a_2\cdots a_n$ from $S$ in~$G$.
If we define a letter-to-letter morphism $h:B^*\to\Sigma^*$
by taking $h((\nabla_i,a))=a$ for all $(\nabla_i,a)\in B$,
then it follows that  $h(L(G')) = L(G) = L$.
{}From $\omega$ the derivation of $w$ from $S$ in $G$ can be reconstructed
deterministically.
In fact, the language $L(G')$ is deterministic context-free.
Hence, there exists a deterministic monotone
RRC-automaton $M$ for this language (see~\cite{JMPV99}).
By interpreting
the symbols of $B$ as auxiliary symbols and by adding the terminal alphabet $\Sigma$,
we obtain a deterministic monotone RRWWC-automaton $M'$ from $M$ such that
$h(L_{C}(M')) =  h(L(M)) = h(L(G')) = L$.
Thus, $(M',h)$ is a deterministic monotone h-RRWWC-automaton with h-proper language~$L$.
Observe that the input language $L(M')$ of $M'$ is actually empty.
\myendproof

Using automata, the construction above illustrates
the linguistic effort to obtain a set of categories (auxiliary symbols)
that ensures the correctness preserving property for the corresponding analysis by reduction.
Note that in its reductions, the automaton $M'$ above only uses delete steps (in a special way).
This is highly reminiscent of the basic (elementary) analysis by reduction learned in (Czech) elementary schools.

{}From our results derived so far we obtain the following separation result,
showing that for deterministic monotone h-RLWW-automata,
the h-proper languages strictly contain the input languages.

\begin{corollary}\label{corCFL}
$\mbox{\sf LRR} = {\mathcal L}({\mbox{\sf det-mon-h-RLWW}})
 \subset {\mathcal L}_{hP}({\mbox{\sf det-mon-h-RLWW }})= \CFL.$
\end{corollary}

In the following we will shortly consider another generalization of the
RLWW-automaton,
the so-called \emph{shrinking} RLWW-automaton, see~\cite{otto147}.
A shrinking RLWW-automaton $M$ is defined just like an RLWW-automaton with the
exception that it is no longer required that each rewrite step of $M$ must be length-reducing.
Instead it is required that there exists a weight function $\omega$ that
assigns a positive integer $\omega(a)$ to each letter $a$ of $M$'s
working alphabet $\Gamma$ such that,
for each cycle-rewriting  $u \Rightarrow_M^c  v$ of~$M$, $\omega(u) > \omega(v)$ holds.
Here the function $\omega$ is extended to a morphism $\omega: \Gamma^* \rightarrow \N$ as usual.
We denote shrinking RLWW-automata by {sRLWW}.
In the following (see the proof of the next proposition)
we will work with a special, linguistically motivated, weight function, which we will call \emph{lexical disambiguation}.
Let us recall that reductions are cycle-rewritings that
decrease the length of the tape contents.

\begin{proposition}\label{shr1}
For each h-RLWW-automaton $M$, there
exists an h-sRLWW-automaton $M_s$ such that $L(M_s) =L_{hP}(M_s) = L_{hP}(M)$,
and there is a one-to-one correspondence between the sets of (length-reducing) reductions of $M_s$ and~$M$.
\end{proposition}

\beginproof
Let  $M=(\hat{M},h) = ((Q, \Sigma,\Gamma, \cent,\$,\delta, q_0,k), h)$ be an h-RLWW-automaton with basic language
$L_C(M)$ and h-proper language $L_{hP}(M) = h(L_C(M))$.
We construct an h-sRLWW automaton
$M_s=(\hat{M}_s,h_s) = ((Q_s, \Sigma,\Gamma_s,\cent,\$,\delta_s,$ $ q_0,k), h_s)$  with weight function $\omega_s$  as follows.
Let $\Gamma_s = \Gamma \cup \{\,\hat{a} \mid a\in \Sigma\,\}$,
where $\hat{a}$ ($a\in\Sigma$) are new working symbols such that $h_s(\hat{a})=a$,
and let $h_s(b)=h(b)$ for all $b\in\Gamma$.

We say that the \emph{degree of lexical ambiguity} of an input symbol $a\in\Sigma$ is~$j$,
if the set $\{\, d \in \Gamma\mid h(d) = a\,\}$ has cardinality~$j$.
This is expressed as ${\rm dga}(a) = j$.
Now we define the weight function $\omega_s$ by taking,
for each $a \in \Sigma$, $\omega_s(a) = {\rm dga}(a) + 1$,
and for each $b \in \Gamma_s \smallsetminus \Sigma$, $\omega_s(b) = 1$.

The automaton $\hat{M}_s$ will work in two phases.
In the first phase, $\hat{M}_s$ nondeterministically
performs the so-called \emph{lexical analysis},
that is, the input symbols on the tape are replaced by symbols from $\Gamma_s \smallsetminus \Sigma$ in such a way
that each $a \in \Sigma$ on the tape is rewritten by a symbol $b$ such that $h_s(b) = a$.
For example, this is achieved by processing the tape from right to left,
replacing a single symbol in each cycle.
Obviously, each of these rewritings is strictly weight-decreasing with respect to the weight function~$\omega_s$.
Phase one ends as soon as the first symbol to the right of the left sentinel $\cent$ has been rewritten.

In the second phase $\hat{M}_s$ simply simulates the given RLWW-automaton~$\hat{M}$
on the rewritten tape contents.
During this phase the new auxiliary symbols $\hat{a}$ are used as substitutes for the input symbols~$a$,
that is, no input symbols occur on the tape during this phase.
Now it follows that $L(M_s) = L_{hP}(M_s) = h(L_C(M)) = L_{hP}(M)$.
As all the rewrite steps of $\hat{M}$ are strictly length-reducing, it follows
that all the simulating steps of $\hat{M}_s$ are shrinking with respect to the weight function~$\omega_s$.
Moreover, it is easily seen that the sets
of reductions of $M_s$ and $M$ coincide (up to the replacement of $a$ by~$\hat{a}$ for all $a\in\Sigma$).
\myendproof

The construction above illustrates the linguistic technique of composing lexical analysis
with the analysis by reduction within a (basic) syntactic analyzer.
Observe, however, that not every language from  ${\calL{\mbox{\sf RLWW}}}$ is the h-proper
language of some h-RLWW-automaton.

\begin{proposition}\label{PGCSL2}
The  language $L_{e}=\{\,a^{2^n}\mid n\in\N \,\} \in {\calL{\mbox{\sf det-RLWW}}} $ is not
the h-proper language of any  h-RLWW-automaton.
\end{proposition}

\beginproof
It is not hard to construct a {det-RLWW}-automaton $M_e$ such that
$L(M_e)=L_{e}$ using the ideas from~\cite{JMPV97}.
In fact, let $M_e=(\{q_0,q_1\},\{a\},\{a,b\},\cent,\$,\delta_e,q_0,3)$ be the deterministic RWW-automaton
that is specified by 
the following transitions:
$$\begin{array}{lcllcl}
\delta_e(q_0,\cent a\$) & = & \Accept, &
\delta_e(q_0,\cent b\$) & = & \Accept, \\
\delta_e(q_0,\cent aa) & = & (q_0,\MVR), &
\delta_e(q_0,\cent bb) & = & (q_0,\MVR),\\
\delta_e(q_0,aaa) & = & (q_0,\MVR), &
\delta_e(q_0,bbb) & = & (q_0,\MVR),\\
\delta_e(q_0,aa\$) & = & (q_1,\mbox{\sf SL}(b\$)),&
\delta_e(q_0,bb\$) & = & (q_1,\mbox{\sf SL}(a\$)),\\
\delta_e(q_0,aab) & = & (q_1,\mbox{\sf SL}(bb)),&
\delta_e(q_0,bba) & = & (q_1,\mbox{\sf SL}(aa)),\\
\delta_e(q_0,\cent ab) & = & \Reject, &
\delta_e(q_0,\cent ba) & = & \Reject,\\
\delta_e(q_1,x) & = & \Restart &\multicolumn{3}{l}{\mbox{for all }x\in \{a,b,\cent,\$\}^{\le 3}.}
\end{array}$$
Given a word $w=a^m$ as input,
$M_e$ will rewrite $w$ from right to left, replacing each factor $aa$ by the symbol~$b$.
This continues until no more rewrites of this form are possible.
If $m$ is uneven, then $M_e$ halts and rejects, otherwise, the word $b^{m_1}$ is obtained with $m_1=\frac{m}{2}$.
In the latter case, $M_e$ will rewrite $b^{m_1}$ from right to left, replacing each factor $bb$ by the symbol~$a$.
Again this continues until no more rewrites of this form are possible.
If $m_1$ is uneven, then $M_2$ halts and rejects, otherwise, the word $a^{m_2}$ is obtained with
$m_2=\frac{m_1}{2}=\frac{m}{4}$.
It can now be easily seen that $L(M_e)=L_{e}$.

To show that  $L_{e} \notin  {\mathcal L}_{hP}({\mbox{\sf h-RLWW}})$,
assume that $L_{e} = L_{hP}(M)$ for some h-RLWW-automaton
$(M,h) = ((Q,\{a\},\Gamma,\cent,\$,\delta,q_0,k),h)$.
Let $z = a^{2^n}\in L_{e}$, where $n$ is a sufficiently large integer satisfying
$2^n-k >  2^{n-1}$, and such that $M$ makes at least one cycle within an accepting computation
on some word $w\in\Gamma^*$ satisfying $w\in L_{C}(M)$ and $h(w) = z$.
It is easily seen that such a word $w$ exists, since otherwise the set of words accepted by $M$ would be a regular language.
The accepting computation on $w$ begins with a reduction of the form
$w  \Rightarrow_M^c   w'$.
Since $w  \Rightarrow_M^c   w'$ is a part of an accepting computation, it follows that $w'\in L_{\rm C}(M)$,
which in turn implies that $h(w')\in L_{e}$.
Thus, $h(w') = a^m$ for some integer $m$ satisfying $2^n-k\le m < 2^n$.
As $2^{n-1} < 2^n-k \le m < 2^n$, this is a contradiction.
Hence, $L_{e}$ is not the h-proper language of any {h-RLWW-automaton}.
\myendproof

Together with Proposition~\ref{shr1} this yields the following proper inclusion.

\begin{corollary}\label{cr1.1}
 $ {\mathcal L}_{hP}({\mbox{\sf h-RLWW}}) \subset {\mathcal L}({\mbox{\sf sRLWW}}).$
\end{corollary}

It can be shown, however, that the language $L_{e}$, which is not the h-proper language
of any h-RLWW-automaton, is in fact the h-pro\-per language
of a deterministic h-sRLWW-automaton that only has length-preserving rewrite steps.

\section{Hierarchies based on window size}\label{seclookahead}
In this section we transfer, extend, and generalize the results of~\cite{fero0}
from input languages to h-proper languages and from RRW-automata to h-RLWW-automata.
We will show that the classes of h-proper languages that are accepted
by  h-RLWW-automata (and by several subclasses of h-RLWW-automata) form infinite ascending
hierarchies with respect to the size of the read/write window.
As separating witness languages, we can actually use the same languages as in~\cite{fero0}.
In addition, we obtain a proper infinite ascending hierarchy within
the left-to-right regular languages that converges to the class LRR, and we obtain
a proper infinite ascending hierarchy within  the context-free languages
that converges to the complete class~CFL.
For input languages of RRWW- and RLWW-automata,  an analogous result is impossible,
as it follows from \cite{fero0,Schluter} that the corresponding hierarchies for the input languages
of RRWW- and RLWW-automata  (and of several subclasses of RLWW-automata)  collapse into only two classes:
those that are accepted by RLWW-automata with window size one, and those that are accepted by RLWW-automata
with window size two or larger.
The new result on the hierarchy for the complete class CFL
stresses the meaning of  the main results from the previous
section, since it is based on them.

Recall that an RWW-automaton must restart immediately after executing a rewrite step (see Subsection~\ref{s4}).
{}From~\cite{fero0} we know that RWW-automata with a window of size one are fairly weak.
Observe that the rewrite steps of any RLWW-automaton with window size one do just delete single symbols.
Accordingly, we have the following characterization, where
REG denotes the class of regular languages.

\begin{lemma}\label{leRWW(1)=Reg}\hspace{\fill}\\[+0.1cm]
$\begin{array}[t]{ccccccccccc}
\REG & = & \calL{\mbox{\sf 1-RC}} & = & \calL{\mbox{\sf 1-R}}& = & \calL{\mbox{\sf 1-RW}} &
= & \calL{\mbox{\sf 1-RWW}}
&=& \calLP{\mbox{\sf 1-h-RWW}}.
\end{array}$
\end{lemma}

On the other hand, RRWW-automata with a window of size one
are more expressive --
already a {1-RR}-automaton can accept a non-context-free language~\cite{fero0}.

Let $D_1$ denote the Dyck language over
the alphabet $\{a_1,\bar{a}_{1}\}$, that is, $D_1$ is
the language that is generated by the context-free grammar
$G=(\{S\}, \{a_1,\bar{a}_1\},S,P)$ with the set
of rules $P= \{S \rightarrow a_1 S \bar{a}_1, S\rightarrow SS, S\rightarrow \lambda\}$.
An alternative way to describe $D_1$ is to interpret $a_1$ as a left bracket
and $\bar{a}_1$ as a right bracket, and then $D_1$ is
the set of all words consisting of well-balanced brackets.
The language $D_1$ is deterministic
context-free, but it is not regular.
However, it is accepted by a
deterministic {2-det-mon-RC}-automaton
that just scans its tape from left to right and deletes the first factor $a_1\bar{a}_1$ that it encounters.

\begin{lemma}\label{leD1indetmonR}{\rm \cite{fero0}}
$D_1 \in \calL{\mbox{\sf 2-det-mon-RC}}.$
\end{lemma}

The next technical lemma lays the foundation for the hierarchies mentioned above.

\begin{lemma}\label{cldetmonRRRWk}
{For all $k \ge 1$,}
${\mathcal L}({\mbox{\sf $(k+1)$-det-mon-RC}}) \smallsetminus {\mathcal L}_{hP}({\mbox{\sf $k$-h-RLWW}})
\not= \emptyset$. 
\end{lemma}

\beginproof
We provide a sequence of languages $\{L_k\}_{k=1}^{\infty}$
which satisfy $$L_k \in {\mathcal L}({\mbox{\sf $(k+1)$-det-mon-RC}})
\smallsetminus{\mathcal L}_{hP}({\mbox{\sf $k$-h-RLWW}}).$$
\begin{enumerate}
\item
For $k=1$, we take the language $L_1=D_1$.
By Lemma \ref{leD1indetmonR}, $D_1\in {\mathcal L}({\mbox{\sf 2-det-mon-RC}})$.
Now assume that $D_1$ is the h-proper language of an
{h-RLWW}-auto\-maton~$M_1=((Q,\Sigma,\Gamma,\cent,\$,q_0,1,\delta),h)$ with window size~one.
As $D_1$ is not regular, there exists an integer $n$ such that
{there is}
a word {$w\in L_C(M_1)$ such that} $h(w) = a_1^n\bar{a}_{1}^n\in D_1$
{and $w$} is not accepted by {a} tail computation of~$M_1$.
Now we consider an accepting computation of $M_1$ {that begins with the restarting configuration $q_0\cent w\$$.}
As this is not a tail computation, it contains a first cycle in which the word $w$
is shortened to a word~$w'$.
As $M_1$ has window size one, this means that $|w'| = 2n-1=|h(w')|$.
Then $h(w')\in L_{hP}(M_1)$, but as $D_1$ does not contain any words of uneven length,
this is a contradiction.
Thus, $D_1$ is not the h-proper language of any {1-h-RLWW}-automaton.

\item
For $k\ge 2$, let $L_k=\{\,a^nc^{k-1}b^n \mid n\ge 0\,\}$.
One can easily design a det-mon-RC-automaton $M_k'$ with a window of size $k+1$
such that $L_k=L(M_k')$,
which shows that $L_k\in {\mathcal L}({\mbox{\sf $(k+1)$-det-mon-RC}})$.
Now assume that $L_k$ is the h-proper language of an
h-RLWW-auto\-ma\-ton $M_k=((Q,\Sigma,\Gamma,\cent,\$,q_0,k,\delta),h)$ with window size~$k$.
Again, as $L_k$ is not regular, there exists an integer $n$ such that
there is  a word $w\in L_C(M_k)$ such that  $h(w) = a^nc^{k-1}b^n\in L_k$
and $w$ is not accepted by a tail computation of~$M_k$.
Now we consider an accepting computation of $M_k$ that begins with the restarting configuration $q_0\cent w\$$.
As this is not a tail computation, it contains a first cycle in which the word $w$
is reduced to a shorter word $w'$ such that $2n-1\le |w'| = |h(w')| \le 2n+k-2$.
However, as the window size of $M_k$ is only~$k$,
we see that either the prefix corresponding to~$a^n$ or the suffix corresponding to~$b^n$
of the word $w$ is not altered through this rewrite step,
which implies that $h(w')$ either has the prefix~$a^n$ or the suffix~$b^n$.
This is a contradiction, as no word from~$L_k$ of length at most $2n+k-2$
has prefix~$a^n$ or suffix~$b^n$.
Thus, it follows that $L_k$ is not the h-proper language of any h-RLWW-automaton with window size~$k$.
\myendproof
\end{enumerate}

Let us note that in the above proof the language $L'_1=\{\,a^nb^n \mid n\ge 0\,\}$
cannot be used to separate ${\mathcal L}(\mbox{\sf 1-R})$ from
${\mathcal L}(\mbox{\sf 2-R})$ or
${\mathcal L}(\mbox{\sf 1-RW})$ from ${\mathcal L}(\mbox{\sf 2-RW})$, etc.,
since $L'_1 \not\in {\mathcal L}(\mbox{\sf 2-RW})$~\cite{fero0}.
The above lemma yields the following hierarchy results.

\begin{corollary}\label{cohierarchies}
For all  $X,Y \in \{\RC,\R,\RW,\hRWW, \hRWWC, \hRWWD,\RRC,\RR,
\RRW,\hRRWW,$ $\hRRWWC,\hRRWWD,  \RLC,\RL,\RLW,\hRLWW, \hRLWWC, \hRLWWD\}$,
all prefixes $\mbox{\it pref}_X, \mbox{\it pref}_Y \in
\{\lambda,{\sf det},{\sf mon},$ $\mbox{\sf det-mon}\}$, and all $k\ge 1$, the following hold:\\[+0.1cm]
$\begin{array}{clcl}
{\rm (a)} & {\mathcal L}_{hP}({\mbox{\sf $k$-pref$_X$-}X}) & \subset &
                      {\mathcal L}_{hP}({\mbox{\sf $(k+1)$-pref$_X$-}X}). \\[+0.1cm]
{\rm (b)} & {\mathcal L}_{hP}({\mbox{\sf $(k+1)$-pref$_X$-}X}) & \smallsetminus &
                      {\mathcal L}_{hP}({\mbox{\sf $k$-pref$_Y$-}Y})
\not= \emptyset.
\end{array}$
\end{corollary}

For example, if $\mbox{\it pref$_X$}=mon$ and $X=\R$, the
expression $\mbox{\it pref$_X$-}X$ denotes $\mbox{mon-R}$,
and for $\mbox{\it pref$_X$}=\lambda$ and $X=\RRW$, the expression $\mbox{\it pref$_X$-}X$
denotes $\RRW$.

{}From Theorem~\ref{m2b} and  Corollary~\ref{cohierarchies},
we see that the classes of h-proper (and input) languages that are accepted by the deterministic monotone
versions of RLC-, RL-, and RLW-automata form infinite strictly ascending hierarchies within the
language class {LRR}.

\begin{corollary}\label{properdetmonhierarchies}
For all ${\sf X} \in \{\RLC,\RL,\RLW\}$,
the following hold:\\[+0.1cm]
$\begin{array}{cl}
{\rm (a)}& \mbox{For all }k\ge 1, \quad{\mathcal L}_{hP}({\mbox{\sf $k$-det-mon-}X}) =
                   {\mathcal L}(\mbox{\sf $k$-det-mon-X}) \subset 
                   {\mathcal L}_{hP}(\mbox{\sf $(k+1)$-det-mon-X}) \subset \mbox{\sf LRR}.\\[+0.1cm]
{\rm (b)} & \bigcup_{k=1}^{\infty} {\mathcal L}_{hP}(\mbox{\sf $k$-det-mon-X}) 
=  \bigcup_{k=1}^{\infty} {\mathcal L}(\mbox{\sf $k$-det-mon-X})   =  \mbox{\sf LRR}.
\end{array}$
\end{corollary}

{}From Theorem~\ref{LPandCFL} and Corollary~\ref{cohierarchies},
we obtain  that the classes of
h-proper languages that are accepted by the monotone versions of deterministic and nondeterministic h-RRWW- and
h-RLWW-automata form proper ascending hierarchies within the class CFL.

\begin{corollary}\label{propermonhierarchies}
For all ${\sf X} \in \{\hRRWW,\hRLWW\}$ and all $\mbox{\it pref} \in\{ \lambda,$ $det\}$ the
following hold:\\[+0.1cm]
$\begin{array}{clclcl}
{\rm (a)} & \mbox{For all }k\ge 1, \quad {\mathcal L}_{hP}(\mbox{\sf $k$-pref-mon-X}) & \subset & {\mathcal L}_{hP}(\mbox{\sf $(k+1)$-pref-mon-X}) & \subset & \CFL .\\[+0.1cm]
{\rm (b)} & \bigcup_{k=1}^{\infty} {\mathcal L}_{hP}(\mbox{\sf $k$-pref-mon-X}) &  = & \CFL.
\end{array}$
\end{corollary}

\section{Restarting automata with multiple {rewrites}}\label{Secmultp}

In this section we consider still another generalization of the
RLWW-automaton,
the h-RLWW-automaton \emph{with multiple rewrites} (mrRLWW-automaton for short).
The aim of this generalization is to obtain a transparent tool which is strong enough to
cover the (surface) syntax of natural languages.
An mrRLWW-automaton $M$ is defined just like an h-RLWW-automaton with the
exception that it is required that, in each  cycle  of a computation, the automaton $M$
must execute a positive number of {\sf SL}-steps.
By ${\sf mrRLWW}(j)$ we denote the class of mrRLWW-automata for which
the number of {\sf SL}-steps in a cycle is limited by the number~$j$.
For mrRLWW-automata we will consider
similar subclasses as for h-RLWW-automata and we will use similar denotations for them.
Further, we use the same important notions for them as for h-RLWW-automata.
\vspace{+0.2cm}

It is easily seen that for deterministic mrRLWW-automata and their basic languages,
the complete correctness (error) preserving property {does not hold}.
However, we have the following useful facts {on their computations}.

\begin{fact}
{\bf (Cycle Error Preserving Property)} \\
Let $M$ be an mrRLWW{-automaton}.
If $ u \Rightarrow_M^{c^*}  v$ and $u\notin L_C(M)$, then $v \notin L_C(M)$.
\end{fact}

\begin{fact}
{\bf (Cycle Correctness Preserving Property)} \\
Let $M$ be a deterministic mrRLWW{-automaton}.
If $ u \Rightarrow_M^{c^*}  v$ and $u \in L_C(M)$, then $v \in L_C(M)$.
\end{fact}

These two facts ensure the transparency for computations of mrRLWW-automata and their
basic and h-proper languages.
{}From the results obtained above, we get
the following corollary.

\begin{corollary}
For all ${\sf X} \in \{\sf mr\RRWW, mr\RRWWD, mr\RRWWC, mr\RLWW,  mr\RLWWD,  mr\RLWWC\}$ and all $\mbox{\it pref} \in\{ \lambda,$ $det\}$ the
following hold:\\[+0.1cm]
$\begin{array}{clcl}
{\rm (a)} &  \CFL & \subseteq & {\mathcal L}_{hP}(\mbox{\sf pref-X}).\\[+0.1cm]
{\rm (b)} &  \CFL & \subset & {\mathcal L}(\mbox{\sf mrRRWW}) \subseteq  {\mathcal L}(\mbox{\sf mrRLWW}) .
\end{array}$
\end{corollary}

This corollary ensures the basic request on the power of mrRLWW-automata
and its variants. The following results support the idea that
mrRLWW-automata are strong enough  to
cover the (surface) syntax of natural languages.

\begin{lemma}
For all $j \ge 1$,
${\mathcal L}({\mbox{\sf det-mrRRC$(j+1)$}}) \smallsetminus {\mathcal L}_{hP}({\mbox{\sf mrRLWW$(j)$}})
\not= \emptyset$.
\end{lemma}

\beginproof
We provide a sequence of languages $\{Lm_j\}_{j=1}^{\infty}$
which satisfy $$Lm_j \in {\mathcal L}({\mbox{\sf det-mrRRC$(j+1)$}})
\smallsetminus{\mathcal L}_{hP}({\mbox{\sf mrRLWW$(j)$}}).$$
\begin{enumerate}
\item
For $j=1$, we take the language $Lm_1= \{ucu \mid u \in \{a,b\}^*   \}$.
It is not hard to show that
$Lm_1\in {\mathcal L}({\mbox{\sf det-mrRRC$(2)$}})$.
In fact, this language is accepted by a {\sf det-mrRRC(2)}-automaton $M_1'$
which, in each cycle, simply deletes the first symbol and the first symbol following the symbol~$c$
if these two symbols coincide. In a tail $M_1'$ {just} accepts~$c$.

Now assume that $Lm_1$ is the h-proper language of an
{h-RLWW}-auto\-maton~$M_1=((Q,\Sigma,\Gamma,\cent,\$,q_0,k,\delta),h)$.
As $Lm_1$ is not regular, there exists an integer $n>k$ such that there is a
word $w \in L_C(M_1)$ such that  $h(w) = a^nb^nca^nb^n\in Lm_1$
and $w$ is not accepted by a tail computation of~$M_1$.
Now we consider an accepting computation of $M_1$ that begins with the restarting configuration $q_0\cent w\$$.
As this is not a tail computation, it contains a first cycle in which the word $w$
is shortened to a word~$w'$.
As $M_1$ has window size $k$, and as it can use exactly one SL-step in a cycle,
it follows than $h(w')$
cannot be from~$Lm_1$.
This is a contradiction, since any cycle in an accepting computation is correctness preserving.
Thus, $Lm_1$ is not the h-proper language of any h-RLWW-automaton.

\item
For $j\ge 2$, let $Lm_j=\{\,{(}uc{)}^{j}u \mid u \in \{a,b\}^* \,\}$.
One can easily design a  det-mrRRC({$j+1$})-auto\-ma\-ton~$M_j'$ {for $Lm_j$},
which shows that $Lm_j\in {\mathcal L}({\mbox{\sf det-mrRRC($j+1$){)}}}$.

Now assume that $Lm_j$ is the h-proper language of an
 mrRLWW($j$)-automaton~$M_j=((Q,\Sigma,\Gamma,\cent,\$,q_0,k,\delta),h)$.
Again, as $Lm_j$ is not regular, there exists an integer $n>k$ such that there is a
word $w{\in L_C(M_j)}$ such that $h(w) = (a^nb^nc)^{j}a^nb^n\in Lm_j$
and $w$ is not accepted by a tail computation of~$M_j$.
Now we consider an accepting computation of $M_j$ that begins with the restarting configuration $q_0\cent w\$$.
As this is not a tail computation, it contains a first cycle in which the word~$w$
is reduced to a shorter word~$w'$, where at least one factor~$v$ of $w$ satisfying $h(v) = a^n$ (or $h(v)=b^n$)
is shortened or at least one~$c$ is deleted,
while another factor~$y$ of~$w$ satisfying $h(y) = a^n$ (or $h(y)=b^n$)  remains unchanged.
This, however, means that $h(w') \notin Lm_j$,
a contradiction.
Thus, it follows that $Lm_j$ is not the h-proper language of any mrRLWW($j$)-automaton.
\myendproof
\end{enumerate}

Thus, we obtain the following consequences.

\begin{corollary}\label{cohierarchies2}
For all  $X,Y \in \{{\sf mrRRC},{\sf mrRR},{\sf mrRRW},{\sf mrRRWW},{\sf mrRRWWD},{\sf mrRRWWC},{\sf mrRLC},$
${\sf mrRL},{\sf mrRLW},{\sf mrRLWW},{\sf mrRLWWD},{\sf mrRLWWC}\}$,
all prefixes $\mbox{\it pref}_X, \mbox{\it pref}_Y \in
\{\lambda,{\sf det}\}$, and all $k\ge 1$, the following hold:\\[+0.1cm]
$\begin{array}{clcl}
{\rm (a)} & {\mathcal L}_{hP}({\mbox{\sf pref$_X$-}X(k)}) & \subset &
                      {\mathcal L}_{hP}({\mbox{\sf pref$_X$-}X(k+1)}). \\[+0.1cm]
{\rm (b)} & {\mathcal L}_{hP}({\mbox{\sf pref$_X$-}X(k+1)}) & \smallsetminus &
                      {\mathcal L}_{hP}({\mbox{\sf pref$_Y$-}Y(k)})
\not= \emptyset.
\end{array}$
\end{corollary}

We believe that already  the class of mrRLWWD($2$)-automata is strong enough to model lexicalized (surface) syntax
of natural languages, {that is,} to model their analysis by reduction.
{In the future} we will {investigate}
the relation{ship} of mrRLWWD($2$)-automata to the class of mildly
context-sensitive languages~\cite{JVW91,JS97}.

\section{Conclusion}\label{SecConclusion}
We have introduced the {h-lexicalized} extension of the RLWW-automaton, which
yields a formal environment that is useful for expressing the lexicalized syntax in computational linguistics.
Then we presented
the input, basic, and h-proper languages of these automata, and
we compared the input languages, which are the languages
that are traditionally considered in automata theory,
to the basic and h-proper languages,
which leads to \emph{error preserving computations} for h-RLWW-automata,
and in the deterministic case
it yields \emph{complete correctness preserving computations}.
Based on the result
that each det-mon-RLWW-automaton can be transformed into a det-mon-RLC-automaton
that accepts the same input language,
we obtained a transformation from mon-RLWW-automata that characterize
the class CFL of context-free languages
through their input languages  to det-mon-h-RLWW-automata that characterize the class CFL
through their h-proper languages.
Through this transformation we have obtained automata with the complete correctness preserving property
and infinite ascending hierarchies within the classes LRR and CFL, based on the size of the read/write window.
Finally, we have introduced classes of restarting automata with transparent computing
which are strong enough to model the syntax of natural languages ({like, e.g.,} Czech,  German or English).
For the future we are interested in deriving a characterization of mildly context-sensitive
languages~\cite{JVW91,JS97} by suitable classes of restarting (and list) automata 
that have the correctness preserving property.
\vspace{+0.2cm}

\noindent
{\bf Acknowledgement.} The authors thank their colleague Franti\v{s}ek Mr\'az for many helpful
discussions on the results presented.

\providecommand{\urlalt}[2]{\href{#1}{#2}} 
\providecommand{\doi}[1]{doi:\urlalt{http://dx.doi.org/#1}{#1}}

\end{document}